\documentclass{article}
\usepackage{graphicx}
\usepackage{amsmath}

\usepackage{authblk}
\usepackage{stix}
     \usepackage{setspace}
     \usepackage{xcolor}
\begin{document}

\title{Quantifying the spatial resolution of the maximum a posteriori estimate in linear, rank-deficient, Bayesian hard field tomography}
\author[1]{Johannes Emmert}
\author[1]{Steven Wagner}
\author[2,*]{Kyle J. Daun}
\affil[1]{\small Technical University of Darmstadt, Department of Mechanical Engineering, Reactive Flows and Diagnostics, Otto-Berndt-Str. 3, 64287 Darmstadt, Germany}
\affil[2]{\small Department of Mechanical and Mechatronics Engineering, University of Waterloo, 200 University Avenue West, Waterloo, ON N2L 3G1, Canada}
\affil[*] {kjdaun@uwaterloo.ca}
\maketitle

\begin{abstract}
Image based diagnostics are interpreted  in the context of spatial resolution. The same is true for tomographic image reconstruction. Current  empirically driven approaches to quantify spatial resolution \cite{Tsekenis.2015} rely on a deterministic formulation based on point-spread functions which neglect the statistical prior information, that is integral to rank-deficient tomography. We propose a statistical spatial resolution measure based on the covariance of the reconstruction (point estimate) and show that the prior information acts as a lower limit for the spatial resolution. Furthermore, the spatial resolution measure can be employed for designing tomographic systems under consideration of spatial inhomogeneity of spatial resolution.
\end{abstract}

\section{Introduction}
Over the last decades absorption spectroscopic linear hard field tomography of gas phase media has been applied to a multitude of engineering problems, including turbines \cite{Ma.2013,Wood.2015,Wright.05.03.201612.03.2016,Fisher.2020}, piston engines \cite{Carey.2000,Wright.2010,Terzija.2015}, exhaust gas aftertreatment \cite{Stritzke.2017,Deguchi.2012}, and coal combustion \cite{Wang.2020} to reconstruct the spatial distribution of temperatures or concentrations. The tomographic reconstruction of these distributions from several line integrals is generally an inverse problem. Due to the limited number of measurement beams these problems are often rank-deficient, requiring  additional information introduced, for example, by regularization methods like Tikhonov regularization \cite{Daun.2010,Tsekenis.2015,Stritzke.2017, Cai.2017}. A more rigorous approach is given by Bayesian inversion methods, which can reinterpret many classical regularization methods like Tikhonov regularization in a statistical framework. \par 
A frequently arising question concerns the quantification of spatial resolution of the reconstruction, be it for comparison to direct imaging methods or to help interpret the reconstruction in terms of resolvable scales in the process. Unlike conventional imaging, the spatial resolution of tomographic reconstructions is not approximately constant but highly inhomogeneous in the imaging domain.  For many non-Bayesian inversion methods the resolution matrix \cite{Bertero.2002,Alumbaugh.2000}, as the name implies, is used as a measure of resolution, since each row of the resolution matrix describes an effective point-spread function (PSF) for one specific location in the imaging plane. As noted by Tsekenis et al. \cite{Tsekenis.2015} the PSF, or its Fourier transform the Modulation Transfer Function (MTF), is a convinient and rigorous way to describe spatial resolution in tomography. In rank-deficient (limited data) tomography, however, not every discrete spatial element is neccessarily captured by a measurement beam as opposed to full-rank tomography like in medical imaging applications for which resolution has been approached with PSFs \cite{Prieto.2010, Nishikido.2008, Thornton.2006}. Hence, the resolution matrix in limited data tomography often contains null PSFs corresponding to "blind spots" in the tomographic beam arrangement. These regions are difficult to interpret in terms of a finite resolution. \par 
Tsekenis et al. \cite{Tsekenis.2015} circumvent this problem by deducing a PSF from the Edge-Spread Functions (ESF) of reconstructed real world phantoms. The underlying assumption hereby is homogeneity and isotropy of the spatial resolution, both being questionable for low beam count measurements. This way, prior information introduced by the regularization is indirectly regarded in the resolution measure and the extensive edge phantoms circumvent the problem of blind spots, yielding heuristic empirical spatial resolution estimates.\par 
In this work we show that, in general, the concept of spatial resolution does not apply to Bayesian tomography which usually gives a finite posterior probability distribution containing arbitrarily large or small spatial frequencies with a certain probability. Instead the posterior probability can be used for uncertainty quantification, yielding credible intervals for the value of each "pixel". Nonetheless, in most real world applications it is common practice to only regard a point estimate, e.g. the maximum a posteriori estimate (MAP), sampled from the posterior distribution. This sampling process accounts to a loss in resolution, which we address in this work.\par 
We begin this paper by reviewing the Bayesian tomography formulation and the deficiencies of the resolution matrix approach in the Bayesian framework. We limit the discussions to a linear hard field tomography problem with normally distributed measurement error and priors formulated as multivariate normal distributions (MVN). In this context we give an expression for the covariance of the MAP estimate and demonstrate its suitability as a resolution measure. Finally, a scalar resolution measure based on a thresholding method proposed by Tsekenis et al. \cite{Tsekenis.2015} is presented.
\section{Tomography}
Linear hard field tomography problems arise from linear integral equations (e.g. Beer-Lambert-Law),  which connect the unknown quantity $f(\chi,\eta)$, which is distributed over two (or more) dimensions $\chi$ and $\eta$, to the measurement (projection) along the $i^\textnormal{th}$ beam, $b_i$,
\begin{equation}
b_i = \int\limits_0^L f(\mathbf{r}(s))\left\lVert\frac{\partial \mathbf{r}}{\partial s}\right\rVert_2\mathrm{d}s,
\label{equ:beerlambert}
\end{equation}
where $L$ is the beam length and $\mathbf{r}(s)$ is the beam path through the measurement volume with parameter $s\in [0,L]$. Discretizing the function $f$ on a linear basis, e.g. a finite element grid with $N$ nodes, yields the representation
\begin{equation}
f(\chi,\eta)\approx \sum\limits_{j=1}^N a_j(\chi,\eta)\mathbf{x}_j = F_\mathbf{x}(\chi,\eta),
\label{equ:FEM_rep}
\end{equation}
where $a_j$ are the finite basis functions and $\mathbf{x}$ the respective weight vector to be solved. The beam integral \eqref{equ:beerlambert} can therefore be approximated by a vector product,
\begin{equation}
b_i = \int\limits_0^L F(\mathbf{r}(s))\left\lVert\frac{\partial \mathbf{r}}{\partial s}\right\rVert_2\mathrm{d}s= \sum\limits_{j=1}^N \mathbf{A}_{i,j}\mathbf{x}_j.
\end{equation}
For $M$ beams this discretization can be summarized in a matrix equation $\mathbf{A}\mathbf{x}=\mathbf{b}$, where $\mathbf{A}$ is the $M\times N$ sensitivity matrix defined by the beam arrangement and the grid.\par
For a fixed measurement system (number of beams and beam arrangement), the rank deficiency of $\mathbf{A}$ hence depends on the chosen grid density, so that any tomography problem may become rank-deficient if the grid resolution is sufficiently fine. In fact the definition of a finite grid density constitutes a prior onto itself \cite{Grauer.2017}. Hence, usually the grid density is chosen sufficiently high to ensure the grid element size is well below any structure size of interest. Therefore, the classification of the problem as rank deficient (limited data, sparse) or full rank is a function of the structure sizes to be measured in the context of the beam arrangement.\par 
Here we assume a general limited data tomography problem with $M< N$. Modeling the unknown variables as random variables and accounting for measurement error leads to the measurement model
\begin{equation}
\mathbf{A}\mathbf{\tilde{x}} = \mathbf{\tilde{b}} +\boldsymbol{\tilde{\epsilon}},
\label{equ:meas_model}
\end{equation}
where $\mathbf{A}$ is the beam sensitivity matrix, $\mathbf{\tilde{x}}$ is the random vector describing the sought after distribution, and $\mathbf{\tilde{b}}$ is the random measurement vector incorporating measurement noise $\boldsymbol{\tilde{\epsilon}}$. Throughout this work random variables are marked by a superscript tilde, $\tilde{\dottedsquare}$, while fixed values and realizations of random variables are written without a tilde. \par
The measurement errors $\boldsymbol{\tilde{\epsilon}}$, introduced by, for example, electrical noise or shot noise, are assumed to follow a MVN distribution with mean value $\boldsymbol\mu_\epsilon$ and covariance $\boldsymbol{\Gamma}_\epsilon$. For many common tomographic problems the error can be assumed independent and identically distributed with 
\begin{equation}
\boldsymbol\mu_\epsilon = 0,
\end{equation}
and
\begin{equation}
\boldsymbol{\Gamma}_\epsilon = \sigma_\epsilon^2 \mathbf{I}.
\label{equ:Gamma_epsilon}
\end{equation}
However, the following discussions are not limited to independent identically distributed errors. Model error that mainly arise due to insufficient grid resolution are assumed to be small compared to measurement noise (sufficiently dense grid), and hence are already accounted for by the error term.
For the other random variables different statistical models will be used as explained in the following sections.
\section{Standard solution approach}
In the Bayesian methodology, the data in $b$ and the unknowns in $x$ are treated as random variables that obey probability density functions (PDFs). These PDFs are related by \cite{Kaipio.2005} 
\begin{equation}
p(\mathbf{x}|\mathbf{b}) = \frac{p(\mathbf{b}|\mathbf{x}) p_\mathrm{pr}(\mathbf{x})}{p(\mathbf{b})},
\end{equation}
where the posterior distribution, $p(\mathbf{x}|\mathbf{b})$, is defined by the model likelihood $p(\mathbf{b}|\mathbf{x})$, which defines the likelihood of observing the data for a hypothetical $\mathbf{x}$, and $p_\mathrm{pr}(\mathbf{x})$ defines what is known about $\mathbf{x}$ before the measurement. The evidence, $p(\mathbf{b})$, is a constant, normalizing the posterior distribution to a proper PDF.
For a linear measurement model and MVN measurement noise the likelihood is given by
\begin{equation}
p(\mathbf{b}|\mathbf{x}) = \frac{1}{\sqrt{(2\pi)^M \mathrm{det}(\boldsymbol\Gamma_\epsilon)}}\mathrm{exp}\left[-\frac{1}{2}(\mathbf{A}\mathbf{x}-\mathbf{b})^T\boldsymbol\Gamma_\epsilon^{-1}(\mathbf{A}\mathbf{x}-\mathbf{b})\right].
\end{equation}
The prior PDF can model any expectations on the behavior of $\mathbf{x}$. As we consider a limited data and therefore rank deficient problem the prior is mandatory in order to avoid a degenerate posterior distribution. The prior constitutes therefore an essential part of the measurement system on the same level as the experimental data. 
The priors should minimize information content beyond the general attributes of the field, e.g. one expects a spatially-smooth distribution due to diffusive transport. For convenience a MVN PDF is often chosen as a model
\begin{equation}
p_\mathrm{pr}(\mathbf{x}) = \frac{1}{\sqrt{(2\pi)^N \mathrm{det}(\boldsymbol\Gamma_\mathrm{pr})}}\mathrm{exp}\left[-\frac{1}{2}(\boldsymbol\mu_\mathrm{pr}-\mathbf{x})^T\boldsymbol\Gamma_\mathrm{pr}^{-1}(\boldsymbol\mu_\mathrm{pr}-\mathbf{x})\right].
\end{equation}
The MVN prior is defined by the expected mean value, $\boldsymbol\mu_\mathrm{pr}$, and, most importantly, the covariance matrix, $\boldsymbol\Gamma_\mathrm{pr}$. The spatial structure expected by the prior is described by the off-diagonal elements in $\boldsymbol\Gamma_\mathrm{pr}$. For example for turbulent flows it might make sense to define a squared exponential covariance \cite{Grauer.2017,Vecherin.2006,Batchelor.1999}
\begin{equation}
\left(\boldsymbol\Gamma_\mathrm{pr}\right)_{i,j} = \sigma_\mathrm{pr}^2 \mathrm{exp}\left(\frac{(\mathbf{r}_i-\mathbf{r}_j)^T(\mathbf{r}_i-\mathbf{r}_j)}{d_\mathrm{corr}}\right)
\label{equ:sqexp}
\end{equation}
With both the model likelihood and the prior PDF modeled as an MVN distribution it can be shown that the posterior distribution is also a MVN distribution \cite{Kaipio.2005} with mean
\begin{equation}
\boldsymbol\mu_\mathrm{post}= \boldsymbol\Gamma_\mathrm{post}\left(A^\mathrm{T}\boldsymbol\Gamma_\epsilon^{-1}(\mathbf{b}-\boldsymbol\mu_\epsilon)+ \boldsymbol\Gamma_\mathrm{pr}^{-1}\boldsymbol\mu_\mathrm{pr}\right),
\label{equ:mu_post}
\end{equation}
and covariance
\begin{equation}
\boldsymbol\Gamma_\mathrm{post}=\left(\boldsymbol\Gamma_\mathrm{pr}^{-1}+\mathbf{A}^\mathrm{T}\boldsymbol\Gamma_\epsilon^{-1}\mathbf{A}\right)^{-1}.
\label{equ:gamma_post}
\end{equation}
This MVN posterior describes the manifold of possible solutions to the tomographic inversion with their corresponding probabilities, and constitutes the outcome of the Bayesian inference. Though it is not trivial to interpret this result directly, it can, for example, be used to derive credible intervals for the quantity of interest at each grid node. Alternatively it is possible to randomly draw a set of solutions from the posterior distribution to visualize credible solutions to the tomographic problem. Unfortunately, for high dimensional tomographic problems, obtaining a representative set of solutions would require a very large number of draws, making this approach impractical. Many practitioners hence default to giving only the solution with the highest probability density, the maximum a posteriori (MAP) estimate.
\section{Maximum a posteriori and Tikhonov Regularization}
For MVN posterior distributions the MAP is equal to the mean value of the distribution, $\boldsymbol\mu_\mathrm{post}=\boldsymbol\mu_\mathrm{MAP}$, given in Equation \eqref{equ:mu_post}. The MAP can also be defined as the solution to a least squares minimization problem \cite{Kaipio.2005}
\begin{equation}
\mathbf{x}_\mathrm{MAP}=\mathrm{arg}\min\limits_\mathbf{x} \left|\left(\begin{matrix}
\mathbf{L}_\epsilon \mathbf{A}\\ \mathbf{L}_\mathrm{pr}
\end{matrix}\right) \mathbf{x} - \left(\begin{matrix}
\mathbf{L}_\epsilon \mathbf{b}\\
\mathbf{L}_\mathrm{pr} \boldsymbol\mu_\mathrm{pr}
\end{matrix}\right)\right|_2^2,
\label{equ:x_MAP_LSQ}
\end{equation}
with 
\begin{equation}
\mathbf{L}_\epsilon = \mathrm{chol}\left(\boldsymbol\Gamma_\epsilon^{-1}\right) \textnormal{, and } \mathbf{L}_\mathrm{pr} = \mathrm{chol}\left(\boldsymbol\Gamma_\mathrm{pr}^{-1}\right),
\end{equation}
where the Cholesky decomposition can also be replaced by any other valid matrix square root. This form highlights the resemblance between the MAP estimation and classical Tikhonov regularization: by setting $\mathbf{L}_\mathrm{pr}=\gamma\mathbf{L}_\mathrm{Tik}$ and $\boldsymbol\mu_\mathrm{pr}=\mathbf{0}$, with the regularization factor $\gamma$ and the Tikhonov matrix $\mathbf{L}_\mathrm{Tik}$ (for example an identity matrix for zeroth order Tikhonov or an Laplace matrix for second order Tikhonov), the MAP estimation becomes the result of a Tikhonov regularization. From the Bayesian perspective Tikhonov regularization resembles a Bayesian inference with a prior defined by the inverse prior covariance $\boldsymbol\Gamma_\mathrm{pr}^{-1}=\gamma^2\mathbf{L}_\mathrm{Tik}^\mathrm{T}\mathbf{L}_\mathrm{Tik}$ and the prior mean $\boldsymbol\mu_\mathrm{pr}=\mathbf{0}$. Note that for the smoothing operators, like second order Tikhonov regularization, $\boldsymbol\Gamma_\mathrm{pr}$ does not exist as $\boldsymbol\Gamma_\mathrm{pr}^{-1}$ is rank deficient. However, this does not limit the applicability of Equation $\eqref{equ:mu_post}$ as it depends on the inverse of the prior covariance.\par 
In summary, Tikhonov regularization amounts to determining a MAP after Bayesian inference, but neglects the actual derivation of a posterior PDF describing the full resolution manifold. Nonetheless, this single sample from the posterior PDF is often given as the convenient solution to the inference problem. With regard to spatial resolution this leads to the definition of the resolution matrix.

\section{Resolution and Point-Spread-Function}
The problem of defining resolution in tomography has been addressed for non-Bayesian methods. Definitions based on the Point-Spread-Function (PSF) seem reasonable and have already been adopted to tomography in a framework based on the resolution matrix \cite{Alumbaugh.2000,DayLewis.2004} as well as in practical approaches \cite{Tsekenis.2015,Yu.2017,Yu.2018}. The interpretation of the PSF as a pulse response of the imaging system can be transferred to tomographic imaging when viewing the tomographic inversion as a black box. The PSF is visualized by applying a initial pulse input to the system in the form of a vector $\mathbf{x}_{\mathrm{pulse},j}$ consisting of all zeros except for the $j^\mathrm{th}$ element being unity. The noise free virtual measurements resulting from this pulse are
\begin{equation}
\mathbf{b}_{\mathrm{pulse},j} = \mathbf{A}\mathbf{x}_{\mathrm{pulse},j}.
\label{equ:res_mat_step1}
\end{equation}
Applying the augmented pseudo inverse \cite{Twynstra.2012},
\begin{equation}
\mathbf{A}^{\#} = \left[
\left(\begin{matrix}
\mathbf{L}_\epsilon \mathbf{A}\\ \mathbf{L}_\mathrm{pr}
\end{matrix}\right)^\mathrm{T}\left(\begin{matrix}
\mathbf{L}_\epsilon \mathbf{A}\\ \mathbf{L}_\mathrm{pr}
\end{matrix}\right)\right]^{-1}\left(\begin{matrix}
\mathbf{L}_\epsilon \mathbf{A}\\ \mathbf{L}_\mathrm{pr}
\end{matrix}\right)^\mathrm{T}\left(\begin{matrix}
\mathbf{L}_\epsilon \\ \mathbf{0}
\end{matrix}\right),
\end{equation}
under the assumption of Tikhonov regularization yields the MAP estimate which is also the PSF of the $j^\mathrm{th}$ point in the grid,
\begin{equation}
\mathbf{x}_{\mathrm{PSF},j} = \mathbf{A}^\#\mathbf{b}_{\mathrm{pulse},j} = \mathbf{A}^\#\mathbf{A}\mathbf{x}_{\mathrm{pulse},j} = \mathbf{R}\mathbf{x}_{\mathrm{pulse},j},
\label{equ:res_mat_def}
\end{equation}
where $\mathbf{R}=\mathbf{A}^\#\mathbf{A}$ is the resolution matrix of the inversion. As indicated by Equation \eqref{equ:res_mat_def} each column $j$ of the resolution matrix defines the PSF of the $j^\mathrm{th}$ grid point.\par
\begin{figure}[h!]
\includegraphics[width=\textwidth]{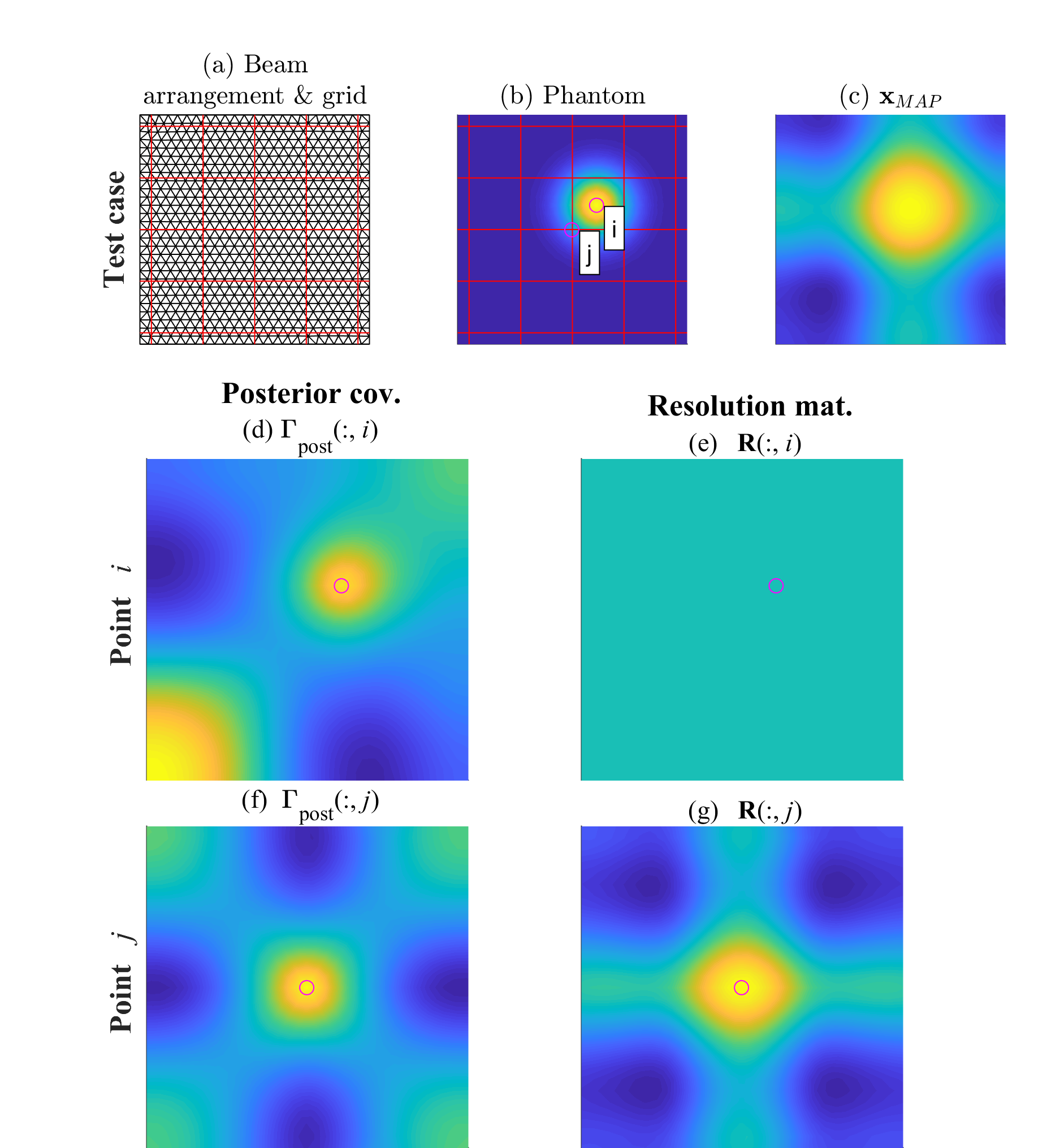}
\caption{Demonstration of posterior covariance and resolution matrix on a Tikhonov regularized test case with beam array and grid (a), phantom (b) and resulting MAP (c). Posterior covariance and resolution matrix are shown for two locations marked with magenta circles: Point $i$ is not traversed by any measurement beam, and Point $j$ is traversed by measurement beams.}
\label{fig:Tikh_res}
\end{figure}

Problems arise when defining PSFs for sparse/limited data tomography: many grid points are not traversed by a single ray, resulting in an null PSF for the corresponding grid point. This scenario is illustrated in Figure \ref{fig:Tikh_res}.
In order to ensure that the grid as a prior does not influence the solution, a sufficiently smooth peak at the off-center location $i$ was chosen as a ground truth for this test case. The beam arrangement consists of only two perpendicular projections, each with five infinitely narrow parallel measurement beams. Many of the grid elements, e.g. point $i$, do not influence the measurement data in our measurement model $\mathbf{A}\mathbf{x}=\mathbf{b}$, and are therefore not probed in the measurement. The measurement uncertainty, $\sigma_\epsilon$, is chosen to be 1~\% of the noise-free peak measurement value, the regularization parameter is set to $\gamma=1$, and $\mathbf{L}_\mathrm{Tik}$ is a Laplace matrix that penalizes curvature. The resulting MAP estimate shows the expected blurring as well as the distortions due to the coarse beam arrangements. For point $j$, which is intersected by two measurement beams, the resolution matrix gives a reasonable result that explains both blurring and distortions. For point $i$ the resolution matrix gives an null PSF (Figure \ref{fig:Tikh_res}(g)).\par
The interpretation of these zero-PSFs or blind spots is not trivial, as there is no direct analogue in conventional imaging, e.g. amplifier and bus real estate on camera chips are usually smaller than the optical blurring kernel of the camera. Also, the interpretation that null PSF means that only infinitely large structures can be resolved, is faulty as structures that span the space between neighboring beams can be resolved.\par
Tsekenis et al. \cite{Tsekenis.2015} avoid these issues by employing the Edge-Spread-Function (ESF) to  first deduce the line-spread function (LSF) and then the PSF. The large phantom structure (two constant value levels divided by a straight line) used to measure the ESF ensures that some information is preserved in the tomographic measurements and the blurring introduced by the regularization ensures that the edge can be resolved in the blind spots of the beam array. Nonetheless, the commonly used direct relation between ESF and LSF implies that the PSF is the same for each grid point (homogeneity). In tomography, however, this is not necessarily true because of varying beam array density, varying quality of the prior, etc. The PSF is then derived from the LSF with the assumption that the LSF represents a cut through the PSF. While this is a crude approximation, exact analytical relations between PSF and LSF either require rotational symmetry and homogeneity of the PSF \cite{Marchand.1964} or for all possible LSF orientations to be measured \cite{Marchand.1965}. The validity of the recovered PSFs is hence disputable. Therefore, the standard definition of resolution does not apply to classical regularized, limited data tomography. These issues can partially resolved within the statistical Bayesian framework.
\section{Resolution in Bayesian Tomography}
Bayesian inference determines a posterior PDF describing the whole manifold of solutions as opposed to a single point estimate. Hence, this solution manifold also contains solutions with a very high spatial frequency content corresponding to small structures. The only lower limit in the possible structure sizes is given by the prior distribution, although this is not a hard limit for Gaussian priors as the probability for very high spatial frequencies is small but finite. Hence, as long as the prior reflects real knowledge about the measurement volume and is consistent with the maximum entropy principle, we can assume that this influence by the prior does not distort the reconstructed image. This gives rise to two possible interpretations of the posterior distribution: either the tomography system resolves all possible fluctuations and therefore has perfect spatial resolution, or it resolves only the  possible fluctuations and therefore has a resolution solely limited by the prior. The more conservative conclusion is that the concept of spatial resolution is not transferable to a manifold or PDF of solutions as it simply describes the current state of knowledge of the measurement volume.\par
This equally unsatisfactory conclusion only concerns the resolution of the posterior distribution. As previously explained it is common practice to depict only a single point estimate from the posterior PDF, for which the definition of spatial resolution again makes sense. However, it should be emphasized that the actual posterior distribution holds the most information and should be used instead of a point estimate with a resolution and uncertainty measure, wherever possible. 
\section{Resolution of the MAP estimate}
Within the Bayesian framework it is convenient to define the resolution of the MAP estimate based on statistical quantities like covariance, matrices instead of PSFs. However, there is a strong connection between PSFs and covariance, as has been already suggested in a non-Bayesian context for tomograms in geostatistics \cite{DayLewis.2004}. This connection is illustrated in the following thought experiment. \par 
Suppose the tomographic imaging system were replaced with a camera to image the physical distribution $\tilde{\mathbf{x}}$, resulting in the camera image $\tilde{\mathbf{x}}_\mathrm{cam}$. The camera optics and imaging sensor have a limited resolution described by the PSFs in the resolution matrix $\mathbf{R}$, making the measurement model
\begin{equation}
\tilde{\mathbf{x}}_\mathrm{cam}= \mathbf{R}\tilde{\mathbf{x}}.
\label{equ:cam_meas_model}
\end{equation} 
We do not have prior knowledge about the structure sizes or correlation lengths of $\tilde{\mathbf{x}}$, but can estimate only its mean and fluctuation width, making our prior choice an MVN prior with mean $\boldsymbol\mu_\mathrm{pr}$ and covariance $\boldsymbol\Gamma_\mathrm{pr}=\sigma_\mathrm{x}^2\mathbf{I}$. With this prior knowledge and Equation \eqref{equ:cam_meas_model} the covariance of the resulting camera image is given as
\begin{equation}
\boldsymbol\Gamma_\mathrm{cam} = \mathbf{R}\boldsymbol\Gamma_\mathrm{pr}\mathbf{R}^\mathrm{T}= \sigma_\mathrm{x}^2\mathbf{R}\mathbf{R}^\mathrm{T}.
\label{equ:cov_from_PSF}
\end{equation}
This shows that the finite PSF of the camera introduces a correlation between the pixels of the camera image. Note that we do not assume a fixed ground truth $\mathbf{x}$ here, but regard the uncertain random variable $\tilde{\mathbf{x}}$ directly, making this covariance matrix of the camera image a property of the measurement system and not of a single measurement. Reversing this operation, i.e. inferring the PSFs from the camera covariance matrix, is an ill-posed problem in itself as it does not feature a unique solution, so it is not directly possible to derive effective PSFs from these covariance matrices. However, the covariance of the camera image itself can be used to define a resolution. For a tomography system the question arises as to how to define such a covariance for a point estimate without knowledge of the PSFs.\par 
While a common first thought is that the posterior covariance matrix is suitable for this purpose, its meaning differs from the covariance of the camera image given previously: it describes possible deviations of the true physical distribution from the given MAP estimate instead of the possible fluctuations of the MAP itself. This is seen in Figure \ref{fig:Tikh_res}, where the plotted posterior covariance matrix columns of nodes $i$ and $j$ do not show a direct relation to the distortions seen in the reconstructed MAP. The posterior covariance instead shows that there exists a negative correlation between the "arm" artifacts introduced by the low projection count and the peak center for the possible deviations from the MAP. Therefore, while the posterior covariance defines the inferred posterior distribution, it cannot be used to determine the resolution of the MAP estimate.\par 
Instead, in a way that is analogous  to the camera covariance a covariance of the MAP point estimate itself can be defined. We follow a similar scheme to the derivation of the resolution matrix given above, but propagate statistical moments instead of fixed values. Instead of ideal uncorrelated point sources, we employ the statistical properties of $\tilde{\mathbf{x}}$ given by $\boldsymbol\Gamma_\mathrm{pr}$ and $\boldsymbol\mu_\mathrm{pr}$.  \par
\begin{figure}[h!]
\includegraphics[width=\textwidth]{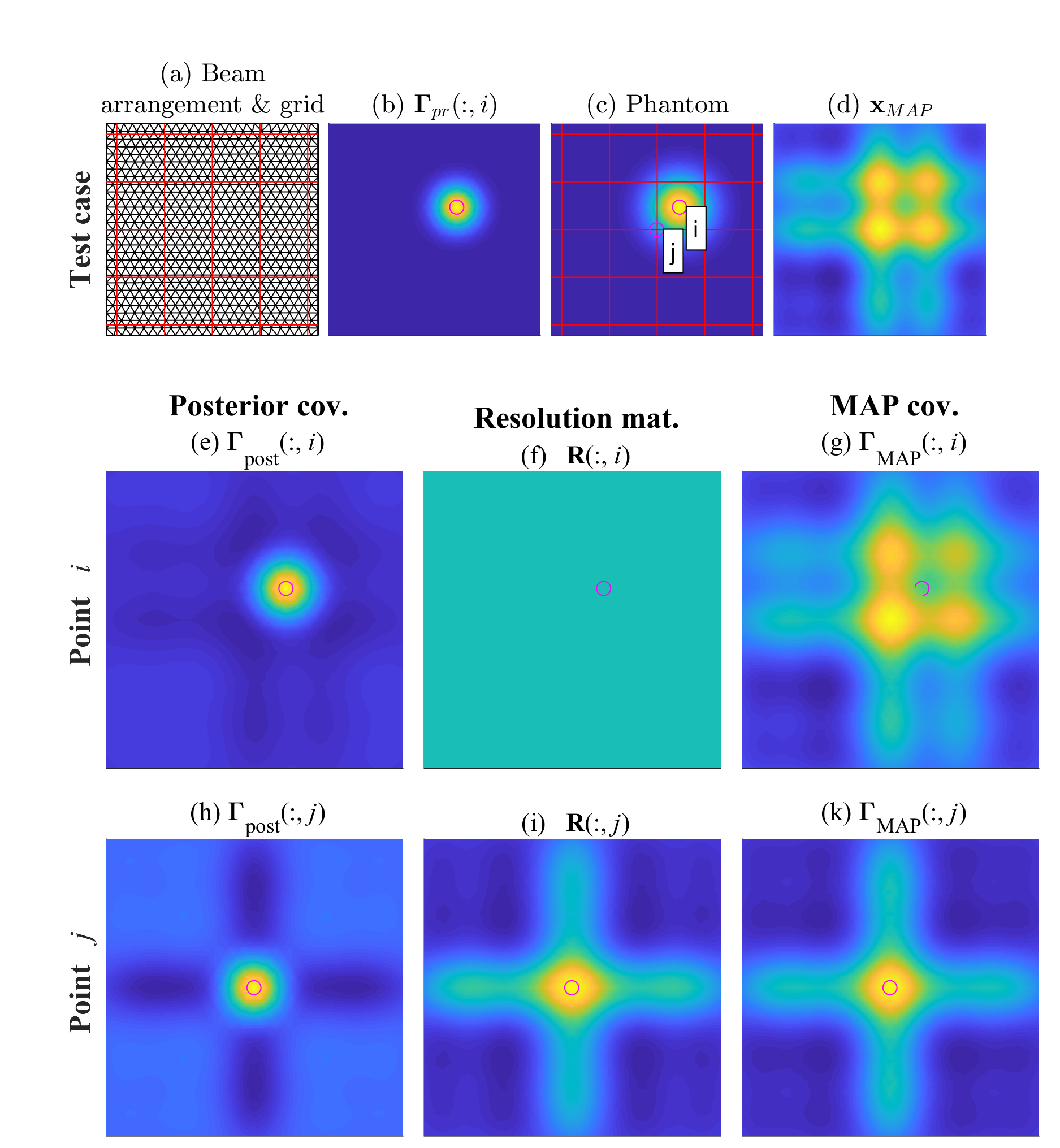}
\caption{Demonstration of posterior covariance, resolution matrix and MAP covariance on a test case with beam array and grid, (a), square exponential prior, (b), phantom, (c), and resulting MAP, (d). Posterior covariance, resolution matrix and MAP covariance are shown for two locations marked with magenta circles: point $i$ is not traversed by any measurement beam, and point $j$ is traversed by measurement beams.}
\label{fig:Sqexp_res}
\end{figure}

 
Given this prior distribution for $\tilde{\mathbf{x}}$ and the distribution of the errors, $\tilde{\boldsymbol\epsilon}$, solving the measurement model in Equation \eqref{equ:meas_model} for  $\tilde{\mathbf{b}}=\mathbf{A}\tilde{\mathbf{x}}-\tilde{\boldsymbol\epsilon}$, allows the prior distribution of measurements  to be determined,
\begin{equation}
p(\mathbf{b}) = \frac{1}{\sqrt{(2\pi)^M \mathrm{det}(\boldsymbol\Gamma_\mathrm{b})}}\mathrm{exp}\left[-\frac{1}{2}(\mathbf{b}-\boldsymbol\mu_\mathrm{b})^T\boldsymbol\Gamma_\mathrm{b}^{-1}(\mathbf{b}-\boldsymbol\mu_\mathrm{b})\right],
\end{equation}
with mean
\begin{equation}
\boldsymbol\mu_\mathrm{b}= \mathbf{A}\boldsymbol\mu_\mathrm{pr} + \boldsymbol\mu_\epsilon,
\end{equation}
and covariance
\begin{equation}
\boldsymbol\Gamma_\mathrm{b}= \mathbf{A}\boldsymbol\Gamma_\mathrm{pr}\mathbf{A}^\mathrm{T} + \boldsymbol\Gamma_\epsilon.
\end{equation}
The calculation of $p(\mathbf{b})$ is analogous to the first step in determining the resolution matrix given in Equation \eqref{equ:res_mat_step1}, but instead of a fixed set of point sources we propagate the statistics of the prior and the measurement error. Accordingly the second step is the propagation of the measurement PDF to the PDF for $\mathbf{x}_\mathrm{MAP}$ based on Equation \eqref{equ:mu_post}, which gives
\begin{equation}
p(\mathbf{x}_\mathrm{MAP}) =\frac{1}{\sqrt{(2\pi)^N \mathrm{det}(\boldsymbol\Gamma_\mathrm{MAP})}}\mathrm{exp}\left[-\frac{1}{2}(\mathbf{x}_\mathrm{MAP}-\boldsymbol\mu_\mathrm{MAP})^T\boldsymbol\Gamma_\mathrm{MAP}^{-1}(\mathbf{x}_\mathrm{MAP}-\boldsymbol\mu_\mathrm{MAP})\right],
\end{equation}
with mean
\begin{equation}
\boldsymbol\mu_\mathrm{MAP}= \boldsymbol\Gamma_\mathrm{post}\left( \mathbf{A}^\mathrm{T}\boldsymbol\Gamma_\epsilon^{-1}(\mathbf{A}\boldsymbol\mu_\mathrm{pr}- \boldsymbol\mu_\epsilon) +\boldsymbol\Gamma_\mathrm{pr}^{-1}\boldsymbol\mu_\mathrm{pr}  \right),
\end{equation}
and covariance
\begin{equation}
\boldsymbol\Gamma_\mathrm{MAP}= \boldsymbol\Gamma_\mathrm{post} \mathbf{A}^\mathrm{T}\boldsymbol\Gamma_\epsilon^{-1}\boldsymbol\Gamma_\mathrm{b}\boldsymbol\Gamma_\epsilon^{-1}\mathbf{A} \boldsymbol\Gamma_\mathrm{post}=\boldsymbol\Gamma_\mathrm{post} \mathbf{A}^\mathrm{T}\boldsymbol\Gamma_\epsilon^{-1}\left(  \mathbf{A}\boldsymbol\Gamma_\mathrm{pr}\mathbf{A}^\mathrm{T} + \boldsymbol\Gamma_\epsilon \right)\boldsymbol\Gamma_\epsilon^{-1}\mathbf{A} \boldsymbol\Gamma_\mathrm{post}.
\label{equ:Gamma_MAP}
\end{equation}
By $p(\mathbf{x}_\mathrm{MAP})$ the statistical properties of the random vector $\tilde{\mathbf{x}}_\mathrm{MAP}$ are defined unconditionally on any concrete realization of $\tilde{\mathbf{x}}$ or $\tilde{\mathbf{b}}$. The MAP covariance, $\boldsymbol\Gamma_\mathrm{MAP}$, hence has a similar meaning to the covariance $\boldsymbol\Gamma_\mathrm{cam}$ given previously and is therefore expected to have a close relation to the theoretical PSFs of the imaging system. This is demonstrated in Figure \ref{fig:Sqexp_res}. 
In this example the same grid, beam arrangement and ground truth were used as for the example in Figure \ref{fig:Tikh_res}, but a squared exponential prior according to Equation \eqref{equ:sqexp} with $d_\mathrm{corr}$ equal to 10$\%$ of the width of the measurement area is applied in place of the Tikhonov prior. This ensures that $\boldsymbol\Gamma_\mathrm{pr}$ exists, enabling the direct application of Equation \eqref{equ:Gamma_MAP}. Priors that do not posses a valid covariance matrix like first and second order Tikhonov priors, will be addressed in the following sections.\par 
Figure \ref{fig:Sqexp_res} demonstrates that while the MAP covariance, $\boldsymbol\Gamma_\mathrm{MAP}$, gives a structurally similar result to the resolution matrix for node $j$, it also gives a correlation structure for node $i$ in between the beams, where the resolution matrix fails (see Figures \ref{fig:Sqexp_res}(g) and \ref{fig:Sqexp_res}(k)). Furthermore, the structure at node $i$ closely resembles the inferred MAP from the ground truth with a peak at $i$. The visible structure of the beam array is an effect of the short correlation lengths not fully spanning the space between measurement beams. This reslt exemplifies how beam arrangement influences resolution.\par 
The MAP covariance does not exhibit the same "blind spots" as the resolution matrix due to the incorporation of the prior information, as long as the prior spans the unprobed grid nodes with a certain correlation. If the prior does not fulfill this requirement the MAP covariance will also feature blind spots similar to those in the resolution matrix. However, if neither prior nor measurement beam span these grid nodes, they are indeed blind (as opposed to the examples given here), either defaulting to the mean value given by the prior or leading to a degenerate posterior and therefor no unique MAP. The prior in limited data tomography is therefore an integral part of the measurement system; this is reflected by the MAP covariance which incorporates the prior information. 
Both the MAP and the MAP covariance resolution can be highly anisotropic in low beam count tomographic imaging. The definition of a scalar resolution quantity is therefore always connected to a further loss in information on the behavior of the imaging system. Nonetheless, in most practical applications a scalar resolution quantity at each location in the tomography domain is preferred.
\section{A resolution measure}
While discussions of the resolution measures based on PSF, optical transfer function (OTF), or modulation transfer function (MTF) can be found throughout optics textbooks \cite{Jahne.2005,Lin.2017}, Tsekenis et al. \cite{Tsekenis.2015} provide a thorough discussion of the intricacies of their application in tomography, concluding that a resolution quantity defined on the MTF or OTF amplitude is more robust than resolution measures defined using the PSF.  Following their definition, the OTF for a node $j$ is given as the Discrete Fourier Transform (DFT) of the PSF,
\begin{equation}
\mathrm{OTF}_j(u,v)=  \mathrm{DFT}(F_{\mathbf{x}_{\mathrm{PSF},j}}(\chi,\eta)),
\end{equation}
where the DFT operates on the two dimensional grid instead of the linear PSF vector itself. The resolution measure in spatial frequency, $f_c$, is then defined by an OTF amplitude threshold relative to the peak OTF of, e.g. $\alpha_\mathrm{th}=20 \%$. The $f_c$ is then transferred to a spatial resolution measure 
\begin{equation}
\delta_j = \frac{1}{2f_{c,j}}.
\end{equation}
As the PSFs are unknown, the OTF cannot be calculated directly, but instead the Fourier transform of the MAP covariance columns can be used:
\begin{equation}
\mathcal{P}_j(u,v)=\mathrm{DFT}(F_{{(\boldsymbol\Gamma_\mathrm{MAP})}_{:,j}}(\chi,\eta)).
\end{equation}
Note that $F_{{(\boldsymbol\Gamma_\mathrm{MAP})}_{:,j}}(\chi,\eta)$ refers to the representation of the spatial distribution on a linear basis, introduced in Equation \eqref{equ:FEM_rep}. The complex Fourier coefficients $\mathcal{P}_j(u,v)$ then describe the complex magnitude of the spatial frequency component $(u,v)$ for the $j^\mathrm{th}$ column of the MAP covariance matrix. Instead of the amplitude of the OTF, we apply the threshold (relative to the maximum) to the amplitude of these Fourier coefficients, $|\mathcal{P}_j(u,v)|$. Therefore the question of choosing the threshold for comparable results to the PSF/OTF based method arises. Note that the units of the amplitudes of $\mathcal{P}_j$ are always the square of the units of the OTF amplitudes. Therefore $\mathcal{P}_j$ resembles a power spectral density of the (non-existent) PSF at node $j$, although this is not a valid identity for a inhomgeneous PSF function. Hence we propose to choose the threshold in $\mathcal{P}$ as the square of the threshold in the OTF,
\begin{equation}
\alpha_{\mathrm{th,\mathcal{P}}} = \alpha_\mathrm{th}^2.
\end{equation} 
Alternatively the square root of the amplitude of $|\mathcal{P}_j|$ can be employed.\par 
As Tsekenis et al. assumed an isotropic PSF their thresholding method gives an unambiguous value. Here we account for the anisotropy of the resolution by calculating, $\boldsymbol\Gamma_\mathrm{MAP}$ and $\mathrm{PSD}$, which are not rotationally symmetric. Thresholding therefore gives a contour line around the zero spatial frequency point in the $\mathrm{PSD}$. In order to attain a conservative scalar estimate of resolution we choose the lowest frequency of the contour line for $f_\mathrm{c} = \sqrt{u_c^2+v_c^2}$.\par  
Before this procedure is demonstrated on a test case, we revisit the treatment of classic Tikhonov smoothness priors with this method. 
\section{Application to improper prior distributions}
While Equations \eqref{equ:mu_post} and \eqref{equ:gamma_post} can be applied to Tikhonov regularization without modifications, the determination of $\boldsymbol\Gamma_\mathrm{MAP}$ according to Equation \eqref{equ:Gamma_MAP} requires a valid prior covariance matrix, $\boldsymbol\Gamma_\mathrm{pr}$, which does not exist in the case of Tikhonov regularization. To get a resolution measure nonetheless, the rank deficiency of the regularization operator (or whitening operator), $\mathbf{L}_\mathrm{Tik}$, needs to be treated. This can be done by taking the singular value decomposition of the $N_\mathrm{Tik}\times N$ operator matrix $\mathbf{L}_\mathrm{Tik}=\mathbf{USV}^\mathrm{T}$, with diagonal elements of $\mathbf{S}$ \cite{Kaipio.2005}
\begin{equation}
s_1 \geq s_2 \geq \dots \geq s_p > s_{p+1} =\dots = s_n=0, \qquad n=\min (N_\mathrm{Tik},N).
\end{equation}
Thus the $(p+1)^\mathrm{th}$ to $n^\mathrm{th }$ column vectors of $\mathbf{V}$ describe the directions in which the prior does not supply any information. These null space basis vectors are summarized as a subspace
\begin{equation}
\mathbf{Q}= \left(v_{p+1},\dots,v_n\right),
\end{equation}
which is used to define the approximate prior covariance \cite{Kaipio.2005}
\begin{equation}
\boldsymbol\Gamma_\mathrm{pr,Tik}=\frac{1}{\gamma^2}\mathbf{L}_\mathrm{Tik}^\dagger\left(\mathbf{L}_\mathrm{Tik}^\dagger\right)^\mathrm{T}+\frac{a^2}{\gamma^2}\mathbf{QQ}^\mathrm{T},
\end{equation}
where $\mathbf{L}_\mathrm{Tik}^\dagger$ is the pseudo inverse of $\mathbf{L}_\mathrm{Tik}$. As $a$ becomes larger this covariance approaches the prior distribution. For common smoothness priors described by Laplace operators or other difference operators, $\mathbf{Q}$ consists of only a single vector with all elements the same value (i.e. the mean value of the distribution is unknown), enabling the discussion of the asymptotic behavior of $\boldsymbol\Gamma_\mathrm{pr,Tik}$. In this case $\mathbf{QQ}^\mathrm{T}$ is proportional to a $N\times N$ matrix of ones, $\frac{1}{c}\mathbf{QQ}^\mathrm{T}=\mathbf{1}_{N\times N}$. Further, note that the absolute magnitude of $\mathcal{P}$ and therefor the magnitude of $\boldsymbol\Gamma_\mathbf{MAP}$ is unimportant as the threshold is chosen relative to  the maximum value. It is hence valid to instead regard an effective MAP covariance $\boldsymbol\Gamma_\mathrm{MAP,eff}=\boldsymbol\Gamma_\mathrm{MAP}/(ca^2) $. For $a\rightarrow \infty$
\begin{eqnarray}
\boldsymbol\Gamma_\mathrm{MAP,eff}= \lim\limits_{a\rightarrow \infty}\frac{1}{ca^2}\boldsymbol\Gamma_\mathrm{MAP}=\\
 = \lim\limits_{a\rightarrow \infty}\frac{1}{ca^2}\boldsymbol\Gamma_\mathrm{post} \mathbf{A}^\mathrm{T}\boldsymbol\Gamma_\epsilon^{-1}\left(  \mathbf{A}\left(   \frac{1}{\gamma^2}\mathbf{L}_\mathrm{Tik}^\dagger\left(\mathbf{L}_\mathrm{Tik}^\dagger\right)^\mathrm{T}+\frac{a^2}{\gamma^2}\mathbf{QQ}^\mathrm{T}             \right)\mathbf{A}^\mathrm{T} + \boldsymbol\Gamma_\epsilon \right)\Gamma_\epsilon^{-1}\mathbf{A} \boldsymbol\Gamma_\mathrm{post}=\nonumber\\
=\frac{1}{\gamma^2} \boldsymbol\Gamma_\mathrm{post} \mathbf{A}^\mathrm{T}\boldsymbol\Gamma_\epsilon^{-1} \mathbf{A}  \mathbf{1}_{N\times N}             \mathbf{A}^\mathrm{T} \Gamma_\epsilon^{-1}\mathbf{A} \boldsymbol\Gamma_\mathrm{post}.\nonumber
\end{eqnarray}
The effective $\boldsymbol\Gamma_\mathrm{MAP,eff}$ can then be used to calculate the PSD and determine the cutoff frequency $f_c$.

\section{Results}
\begin{figure}[h!]
\includegraphics[width=\textwidth]{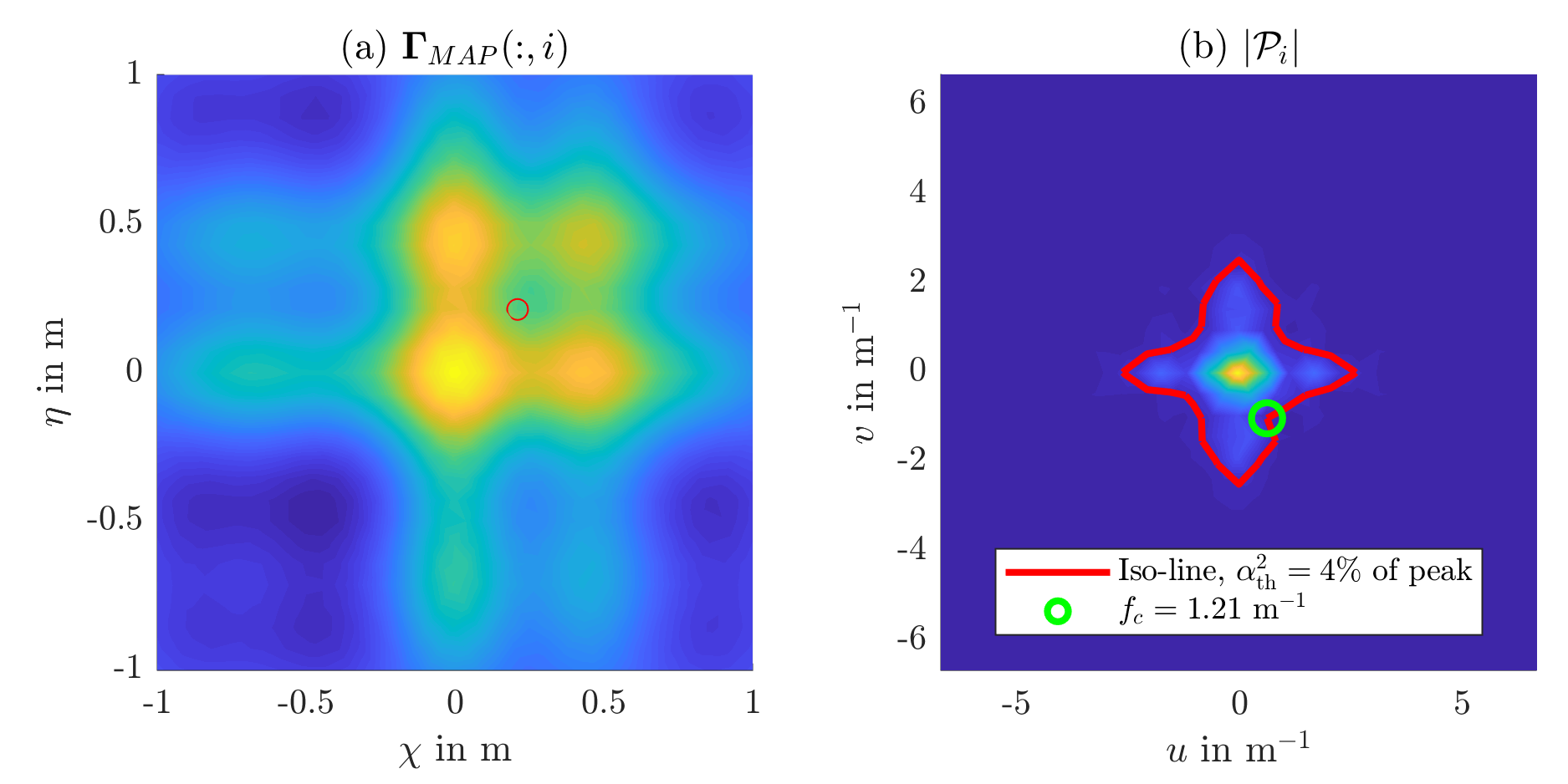}
\caption{Determination of a scalar resolution measure from the MAP covariance.(a) MAP covariance for the squared exponential example in Figure \ref{fig:Sqexp_res} at Point $i$. (b) PSD column for point $i$ with the threshold contour at $\alpha_\mathrm{th}^2=(0.2)^2=4\%$ of peak (following the value used by Tsekenis et al. \cite{Tsekenis.2015}).}
\label{fig:thresh}
\end{figure}

\begin{figure}[h!]
\centering
\includegraphics[scale=0.8]{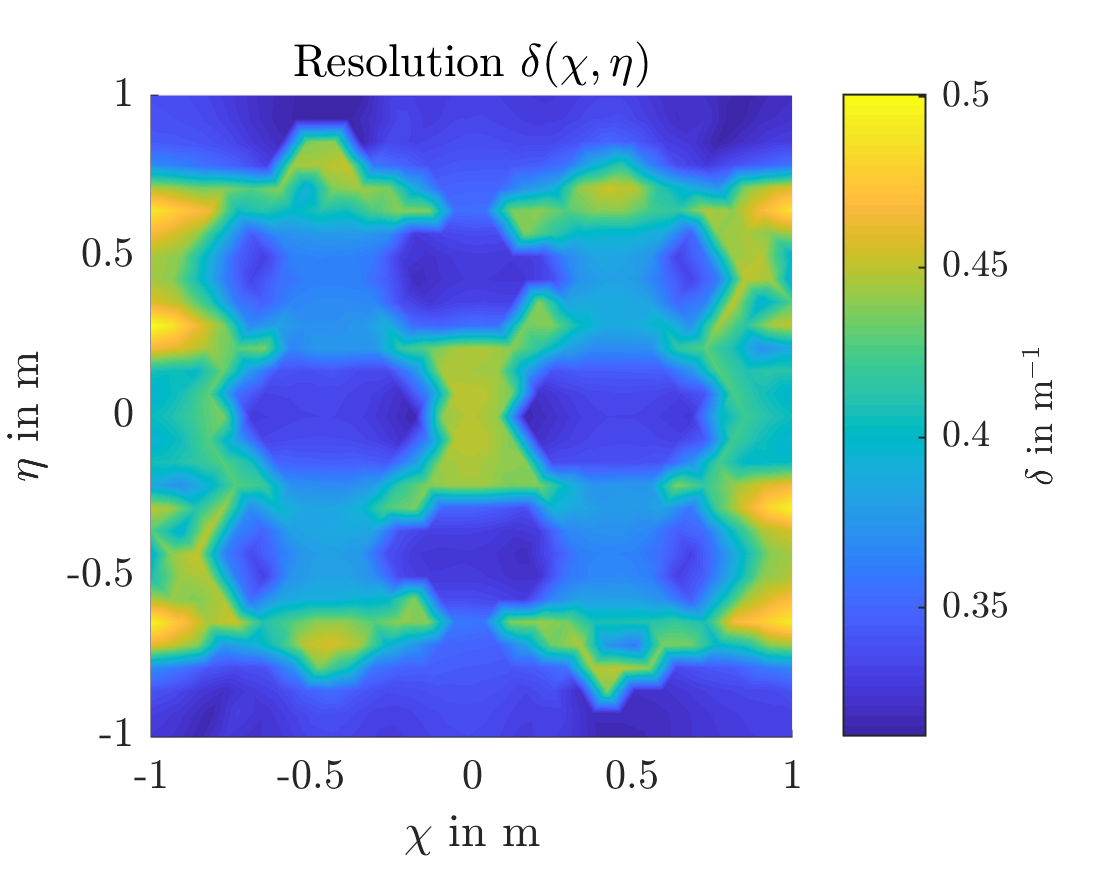}
\caption{Spatial resolution mapped to the reconstruction domain.}
\label{fig:spatialres}
\end{figure}

To demonstrate the determination of a scalar resolution estimate we again employ the example given in Figure \ref{fig:Sqexp_res}. The MAP covariance for point $i$ and the corresponding PSD column are depicted in Figure \ref{fig:thresh}.
As can be seen in Figure \ref{fig:thresh}b the cut off frequency is determined to be $f_c=1.02$~m$^{-1}$, resulting in a scalar resolution measure of $\delta=0.4911$~m. Determining this resolution measure for every grid point yields the resolution map in Figure \ref{fig:spatialres}.

Artifacts originating from the coarse beam array are visible in the resolution pattern, but the detailed influence of the beam count and arrangement, which is introduced by the beam sensitivity matrix, $\mathbf{A}$, in Equation \eqref{equ:Gamma_MAP}, is not apparent.\par
\begin{figure}[h!]
\includegraphics[width=\textwidth]{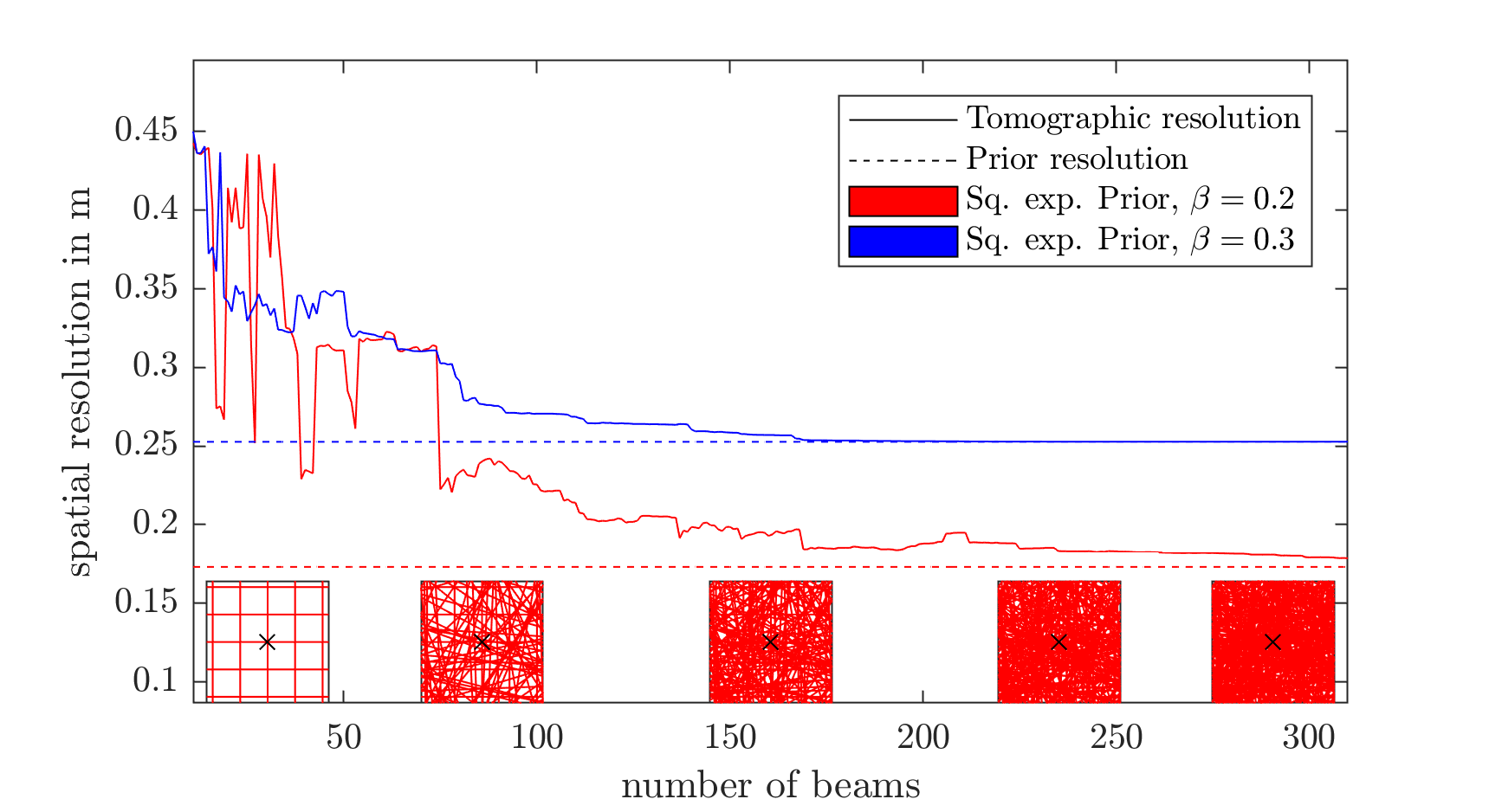}
\caption{Spatial resolution over beam count for two squared exponential priors with different spatial correlation lengths.}
\label{fig:res_over_density}
\end{figure}

 
Instead we again focus on the resolution in a single point of the spatial domain to discuss the influence of the measurement information quantity on resolution. 
In order to investigate the influence of beam array density, randomly oriented beams are added to the orthogonal array.
Figure \ref{fig:res_over_density} shows that as expected resolution improves as the beams are added. 
By increasing the beam count further the spatial resolution at each single point approaches a fixed value given by the prior distribution, which acts as a lower limit of the spatial resolution. This is illustrated in Figure \ref{fig:res_over_density}, which depicts resolutions found using two squared exponential priors, one with a correlation length parameter $\beta=0.2~$m (shorter spatial correlation length) and one with $\beta=0.3~$m (longer spatial correlation length). The prior with the shorter correlation length is the less restrictive prior, providing less support of the solution but allowing for higher spatial resolution. The asymptotic resolution values of 0.173~m for $\beta=0.2~$m and 0.256~m for $\beta=0.3~$m match these expectations. Furthermore, the dashed lines in Figure \ref{fig:res_over_density} represent the prior resolution which is calculated by applying the scalar resolution measure to the prior covariance instead of the MAP covariance. In accordance with the previous explanation the prior resolution represents the asymptote to the tomographic resolution in Figure \ref{fig:res_over_density}. Hence, the prior gives a lower bound for the spatial resolution, and when this lower bound is reached additional measurement beams do not significantly improve resolution. This agrees with studies of the influence of the beam arrangement on resolution using Tsekenis approach \cite{Liu.2019}: the resolution approaches a limiting value, although this could not directly be explained using Tsekenis empirical approach.

\section{Conclusion}
This work discusses spatial resolution measures for linear Bayesian hard field tomography. For the posterior PDF a spatial resolution measure cannot be directly defined due to the arbitrarily large frequency content of candidate solutions. 
Rather, it is only possible to define the resolution of a specific point estimate, in this case the maximum a posteriori estimate.
Especially in the case of rank deficient tomographic problems the classical approach of measuring or calculating the point-spread function suffers from indefiniteness in blind spots that are not traversed by measurement beams. These problems stem from the contradiction between the assumed prior and the assumption of point spread functions, or, put differently, from disregarding the prior knowledge used for inversion. To remedy these problems we propose a statistical measure, the covariance matrix of the maximum a posteriori estimate, instead of point-spread functions. The statistical formulation allows for the incorporation of the prior PDF into the resolution measure. The use of spatial correlations of the MAP allows for a similar treatment as point-spread functions, namely transfer to spatial frequency domain and thresholding, giving a scalar resolution measure for each point in the measurement domain.\par 
This resolution measure is influenced by measurement noise, beam arrangement, and the prior. It constitutes an important mathematically-valid quality criterion that can be employed in the concept phase of the design of a tomographic measurement system. 

\section{Acknowledgments}
The authors thank Prof. Jari Kaipio for the helpful discussions about Bayesian inference and tomographcy in general.

\section{Funding}
Deutsche Forschungsgemeinschaft (Projektnummer 215035359 – TRR 129). \\
Natural Sciences and Engineering Research Council of Canada DG RGPIN-2018-03765.
\bibliography{literature}
\end{document}